\documentclass[a4paper]{article}

\usepackage{epsfig}

\newcommand{\req}[1]{(\ref{#1})}
\newcommand{\Beq}{\begin{equation}}
\newcommand{\Eeq}{\end{equation}}
\newcommand{\beq}{\begin{displaymath}}
\newcommand{\eeq}{\end{displaymath}}
\newcommand{\Beqa}{\begin{eqnarray}}
\newcommand{\Eeqa}{\end{eqnarray}}
\newcommand{\beqa}{\begin{eqnarray*}}
\newcommand{\eeqa}{\end{eqnarray*}}
\newcommand{\nUV}{\mu_{R}^{2}}
\newcommand{\nIR}{\mu_{F}^{2}}
\newcommand{\nO}{\mu_{0}^{2}}
\newcommand{\x}{\overline{x} \,}
\newcommand{\y}{\overline{y} \,}

\begin{document}

\begin{flushright}
IRB-TH-2/99
\end{flushright}

\vspace*{1.7cm}

\begin{center}

\begin{large}

{\bf 
     ON THE COMPLETE NEXT-TO-LEADING ORDER pQCD
       PREDICTION FOR THE PION FORM FACTOR\footnote{
       Presented by K. Passek at Nuclear and Particle Physics
       with CEBAF at Jefferson Lab Conference (Dubrovnik'98)}
}

\end{large}

\end{center}

\begin{center}
        BLA\v{Z}ENKA MELI\'{C},
        BENE NI\v{Z}I\'{C}
        and KORNELIJA PASSEK\footnote{Electronic addresses: 
	                     melic@thphys.irb.hr,
                             nizic@thphys.irb.hr, passek@thphys.irb.hr}
\end{center}

\begin{center}
\begin{small}
{\em Theoretical Physics Division, Rudjer Bo\v{s}kovi\'{c} Institute, 
        P.O. Box 1016, 10001 Zagreb, Croatia}
\end{small}
\end{center}

\begin{center}
\begin{small}
January 1999
\end{small}
\end{center}

\vspace*{0.35cm}
\noindent
%\begin{abstract}
We comment on the results of a complete leading-twist
next-to-leading order QCD analysis of the 
spacelike pion electromagnetic form factor at large-momentum
transfer $Q$.
For the asymptotic distribution amplitude, 
we have examined the sensitivity of the predictions 
to the choice of the renormalization scale.
The results show that reliable 
perturbative predictions for the pion electromagnetic form factor 
can already be made at a momentum transfer $Q$ below $10$ GeV. 
%\end{abstract}
\\

\vspace*{0.25cm}
\noindent
PACS numbers: 13.40.Gp, 12.38.Bx \\[0.25cm]
Keywords: exclusive processes, perturbative QCD, next-to-leading order, 
          pion form factor, distribution amplitudes,
	  renormalization and factorization scales

\section{{\it{\Large Introduction}}}

The framework for analyzing 
exclusive processes at large-momentum transfer 
within the context
of perturbative QCD (pQCD)
has been developed in the late seventies 
\cite{sHSA}.
It was demonstrated, to all orders in perturbation theory,
that exclusive amplitudes involving
large-momentum transfer factorize into a convolution
of a process-independent and perturbatively incalculable
distribution amplitude (DA), 
one for each hadron involved in the amplitude, with
a process-dependent and perturbatively calculable 
hard-scattering amplitude.
In the standard hard-scattering approach (sHSA), hadron is regarded
as consisting only of valence Fock states, transverse quark momenta
are neglected (collinear approximation) as well as quark masses.

Although the pQCD approach of Ref. \cite{sHSA}
undoubtedly represents an adequate and efficient tool for analyzing
exclusive processes at very large momentum transfer,
its applicability to these processes at experimentally accessible 
momentum transfer has been long debated and attracted much attention.
There are several important issues regarding this subject.
Let us mention first that in a moderate energy region (a few GeV) 
soft contributions 
(resulting from the competing, so-called, Feynman mechanism)
can still be substantial \cite{soft}, but the estimation of their size is 
model dependent. 
Further on, the self-consistency of the pQCD  approach was questioned
regarding the nonfactorizing end-point contributions
\cite{IsLS84etc}. 
It has been shown, that the incorporation of 
the Sudakov suppression
in the so-called modified perturbative approach (mHSA) \cite{LiS92etc,JaK93} 
effectively eliminates soft contributions from the end-point regions and 
that the pQCD approach 
to the pion form factor begins to be self-consistent for a momentum transfer
of about $Q^2 > 4$ GeV$^2$ \cite{LiS92etc}. 
However, in the pQCD approach to exclusive processes one still has to check
its self-consistency by studying radiative corrections.

It is well known that, unlike in QED, the 
leading-order (LO) predictions in pQCD 
do not have much predictive power, and that higher-order corrections
are important. 
In general, they have a stabilizing effect 
reducing the dependence of the 
predictions on the schemes and scales. 
Although  the 
LO predictions within the sHSA (as well as, the mHSA) 
have been obtained 
for many exclusive processes,
only a few processes have been 
analyzed at the next-to-leading order (NLO):
the pion electromagnetic form factor 
\cite{pioneff,KaM86,MeN98},
the pion transition form factor \cite{KaM86,piontff} 
(and \cite{MuR97} in mHSA),
and the process 
$\gamma \gamma \rightarrow M \overline{M}$ ($M=\pi$, $K$) \cite{Ni87}.

In our recent work \cite{MeN98} we have clarified 
some discrepancies between
previous results \cite{pioneff}, 
and by including the complete closed form 
for the NLO evolution of the pion DA 
derived recently  \cite{Mu94etc}, 
we have obtained the complete NLO prediction
for the pion electromagnetic form factor. 

In this work we would like to give a short summary of 
our calculation
(for details and notation see \cite{MeN98}) with special
emphasis on the proper choice of the renormalization scale.

\section{{\it{\Large Pion electromagnetic form factor in the sHSA}}}

In leading twist, the pion electromagnetic form factor
can be written as 
\Beq
    F_{\pi}(Q^{2})=\int_{0}^{1} dx \int_{0}^{1} dy \;
                    \mathit{\Phi}^{*}(y,\nIR) \;
                          T_{H}(x,y,Q^{2},\nUV,\nIR) \;
                    \mathit{\Phi}(x,\nIR) \, .
\label{eq:piffcf}
\Eeq

Here $Q^2=-q^2$ is the momentum transfer in the process and
is supposed to be large, 
$\mu_R$ is the renormalization scale,
and $\mu_F$ is the factorization scale at which
soft and hard physics factorize.

The hard-scattering amplitude $T_{H}(x,y,Q^{2},\nUV,\nIR)$ 
is the amplitude for a parallel 
$q_1 \overline{q}_2$ pair of the total momentum $P$, 
with the constituents carrying the momentum fractions $x$ and $\x=1-x$,
hit by a virtual photon $\gamma^*$ of momentum $q$ 
to end up as a parallel
$q_1 \overline{q}_2$ pair of momentum $P'=P+q$,
with the constituents sharing fractions $y$ and $\y=1-y$.
It can be calculated in 
perturbation theory and represented as a
series in the QCD running coupling constant $\alpha_S(\nUV)$:
\Beqa
  \lefteqn{T_{H}(x,y,Q^2,\nUV,\nIR)=
   \alpha_{S}(\nUV) \, T_{H}^{(0)}(x,y,Q^2)}
            \nonumber \\ & & \qquad \qquad \times \;
   \left[ 1 + \frac{\alpha_{S}(\nUV)}{\pi} \, 
	       T_{H}^{(1)}(x,y,\nUV/Q^2,\nIR/Q^2) 
                + \cdots \right] \, .
\label{eq:TH}
\Eeqa
There are, up to the order we are calculating,
4 LO and 62 NLO Feynman diagrams that contribute 
to the $(q_1 \overline{q}_2) + \gamma^* \rightarrow (q_1 \overline{q}_2)$
amplitude.
We have used the dimensional regularization method and the
$\overline{MS}$ renormalization scheme in our calculation 
(details and results are presented in \cite{MeN98}). 

The intuitive interpretation of the pion DA $\mathit{\Phi}(x,\nUV)$
($\mathit{\Phi}^*(y,\nUV)$)
is that it represents a probability amplitude for finding the
valence 
$q_1 \overline{q}_2$ Fock state in the initial (final) pion.
The function ${\mathit\Phi}$ is intrinsically nonperturbative, 
but its evolution can be calculated perturbatively.
It is advantageous to express the DA 
(further on, we use the function $\phi$ normalized to unity)
in terms of the Gegenbauer polynomials 
\Beq
     \phi(x)=\phi(x,\mu_0^2)=\phi_{as}(x) \,
	  \left[ 1 + \sum_{n=2}^{\infty} {}' B_n \:  C_n^{3/2}(2 x - 1) 
	 \right]
	   \, .
\label{eq:phiexp}
\Eeq
Here $\phi_{as}(x)=6 x (1-x)$ is the asymptotic DA which represents
the solution of the DA evolution equation for $\nIR \rightarrow \infty$,
while the coefficients $B_n$ ($n$ even) are obtained 
using some nonperturbative
techniques at energy $\nO$. 
The DA is then evoluted to the energy $\nIR$ and 
the DA evolution up to the NLO has the form 
\Beq
   \phi(x,\nIR)=\phi^{LO}(x,\nIR) + \frac{\alpha_S(\nIR)}{\pi} \,
               \phi^{NLO}(x,\nIR) \, ,
\label{eq:evDA}
\Eeq
where
we turn to \cite{MeN98} for detailed expressions.

Two most exploited choices for the pion DA \req{eq:phiexp}
are $\phi_{as}(x)$ and $\phi_{CZ}(x)$, 
for which $B_2=0$ and $B_2=2/3$, respectively, 
while $B_n=0$ for $n>2$.
Unlike $\phi_{as}(x)$,
the $\phi_{CZ}(x)$ is a strongly end-point concentrated distribution
and its form has been obtained using the method of QCD sum rules
\cite{ChZ84}.
There is no LO evolution for $\phi_{as}(x,\nIR)$ and the NLO evolution is
tiny.
As we have shown in \cite{MeN98}, the inclusion of the LO evolution
is crucial when one tends to obtain meaningful results
with the $\phi_{CZ}$ function, and even the NLO evolution is significant.

Although the numerical results for the pion electromagnetic form
factor obtained by using the $\phi_{CZ}(x,\nIR)$ distribution are higher
and closer to the existing experimental data,
there are compelling theoretical results which disfavor
the $\phi_{CZ}$ distribution: theoretical predictions for the
pion transition form factor are in very good agreement with the data assuming
that the pion distribution amplitude is close to the asymptotic one,
i.e., $\phi_{as}(x,\nIR)$ \cite{Rad95etc}; the estimation of the size of the
soft contributions  to the pion electromagnetic form factor 
\cite{soft} indicates the asymptotic form of the pion DA, and 
even the self-consistency of the derivation of the
$\phi_{CZ}(x,\nIR)$ distribution from the QCD
sum rules was criticized \cite{Rad98}.
Taking this into account, one expects that the pion DA does not differ much
from the $\phi_{as}$ and in this work we comment only on
the results obtained with $\phi_{as}(x,\nIR)$.  

By inserting \req{eq:TH} and \req{eq:evDA} into \req{eq:piffcf}
one obtains the complete NLO pQCD expression for the pion electromagnetic
form factor.
Generally, one can express the NLO form factor as
\Beq
  F_{\pi}(Q^2,\nUV,\nIR) = F_{\pi}^{(0)}(Q^2,\nUV,\nIR)
          + F_{\pi}^{(1)}(Q^2,\nUV,\nIR) \, .
\label{eq:Fpi}
\Eeq
The first term in \req{eq:Fpi} is the LO contribution, while
the second term is the NLO contribution coming
from the NLO correction to the hard-scattering amplitude
as well as arising from the inclusion of the NLO
evolution of the DA. For the results obtained 
using $\phi_{as}(x,\nIR)$ distribution,
the effect of the NLO evolution of the DA is negligible ($\approx 1\%$).

\section{{\it{\Large Choosing the factorization and renormalization scales }}}

The physical pion form factor $F_{\pi}(Q^2)$,
represented at the sufficiently high $Q^2$ by the factorization formula 
\req{eq:piffcf}, is independent of the renormalization scheme and of the
renormalization and factorization scales, $\mu_R$ and $\mu_F$, respectively. 
Truncation of the perturbative series of $F_{\pi}(Q^2)$
at any finite order causes a residual dependence on the scheme as well as
on the scales (which is already denoted in Eq. \req{eq:Fpi}).
We approximate $F_{\pi}(Q^2)$ 
only by  two terms of the perturbative series and
hope that we can minimize higher-order corrections
by  a suitable choice of $\mu_{R}$ and $\mu_{F}$, 
so that the LO term $F_{\pi}^{(0)}(Q^2,\nUV,\nIR)$
gives a good approximation 
to the complete sum $F_{\pi}(Q^2)$.

The simplest and widely used choice for the  $\mu_R$ and $\mu_F$ scales
is
\Beq
      \nUV=\nIR=Q^2 \, ,
\label{eq:nRQ}
\Eeq
the justification for the use of which is mainly pragmatic.
Physically, a more appropriate choice for $\nUV$ 
would be that corresponding to the characteristic
virtualities of the particles in the parton subprocess,
which is considerably lower  than the overall momentum transfer
$Q^2$ (i.e., virtuality of the probing photon).
The physically motivated choices we are using are
\Beqa
    \nUV&=& \x \y Q^2 \, , 
\label{eq:nRg} \\
   \nUV&=&\sqrt{(\x \y Q^2) (\y Q^2)} \, ,
\label{eq:nRgq} \\
    \nUV&=&e^{-5/3} \x \y Q^2 \, . 
\label{eq:nRge}
\Eeqa
These correspond, respectively, to the (LO) gluon virtuality,
geometrical mean of the gluon and quark virtualities
(an attempt to take into account that in QCD, unlike in QED,
the coupling is renormalized not only by the vector particle propagator,
but also by the quark-gluon vertex and the quark-propagator,
Dittes and Radyushkin \cite{pioneff}),
and to the choice of the renormalization scale
according to the Brodsky-Lepage-Mackenzie (BLM) procedure \cite{BrL83}.
The essence of the BLM procedure is that all vacuum-polarization 
effects from the QCD $\beta$ function are resummed into the
running coupling constant, and, 
as a result of the choice \req{eq:nRge}, 
$T_{H}^{(1)}(x,y,\nUV/Q^2,\nIR/Q^2)$ from 
\req{eq:TH} becomes $n_f$  (i.e. $\beta_0$) independent.

A glance at Eq. \req{eq:piffcf}, where 
the coupling constant
$\alpha_S(\nUV)$ appears under the
integral sign, reveals that any of the choices of $\mu_{R}$
given by (\ref{eq:nRg}--\ref{eq:nRge})
leads immediately to the problem
if the usual one-loop formula 
for the effective QCD running coupling constant is employed.
To circumvent this,
one can introduce a cutoff in one-loop formula 
with the aim of preventing the effective coupling from becoming infinite
for small gluon momenta.
There are number of proposals for the form of the coupling constant
$\alpha_S(\nUV)$ for small $\nUV$ \cite{alphaSmod,alphaSmodn}, 
but its implementation in this calculation
demands the more refined treatment (see the discussion in \cite{MeN98}).
Alternatively, one can choose $\nUV$ to be an effective constant
by taking $\nUV=\left< \nUV \right>$.
Hence, in this work 
we have replaced the expressions (\ref{eq:nRg}-\ref{eq:nRge}) by
their respective averages
\Beqa
    \nUV&=&\left< \x \right>^{2} Q^2 \, , 
\label{eq:nRx2} \\
    \nUV&=&\left< \x \right>^{3/2} Q^2 \, ,
\label{eq:nRx3} \\
    \nUV&=&e^{-5/3} \left< \x \right>^{2} Q^2 \, . 
\label{eq:nRx2e}
\Eeqa
Owing to the fact that $\phi_{as}(x,\nIR)$ (as well as any other pion DA)
is centered around the value $x=0.5$, 
we take the average value of the momentum fraction to be
$\left< \x \right>=0.5$.

We have shown in \cite{MeN98} that the results
depend very weakly on the choice of the factorization scale $\mu_F$.
Actually, 
taking $\nIR$ to be an effective constant, i.e.,
$\nIR=\left< \nIR \right>$, 
the only $\nIR$ dependence of
the results obtained using $\phi_{as}(x,\nIR)$ distribution
comes from the NLO evolution of the DA which is negligible.
In the following we take $\nIR=Q^2$.

\section{{\it{\Large Numerical results}}}

We take that a perturbative prediction for $F_{\pi}(Q^2)$ 
can be considered reliable provided 
the corrections to the LO prediction are reasonably small
($< 30\%$) and
the expansion parameter (effective QCD coupling constant) is
acceptably small ($\alpha_S(\mu_R^2)< 0.3$ or $0.5$).
The consistency with the experimental data
is not of much use here since 
reliable experimental data
for the pion form factor exist for $Q^2 \leq 4$ GeV$^2$ \cite{Be78}
 i.e., 
outside the region in which the perturbative treatment based on
Eq. \req{eq:piffcf} is justified.
It should also be mentioned that 
one can find controversial arguments in the literature \cite{exp?} 
regarding the reliability of the analysis of Ref. \cite{Be78}.
The new data in this energy region are expected from the CEBAF experiment
E-93-021.

\begin{figure}
\begin{tabular}{cc}
\epsfig{file=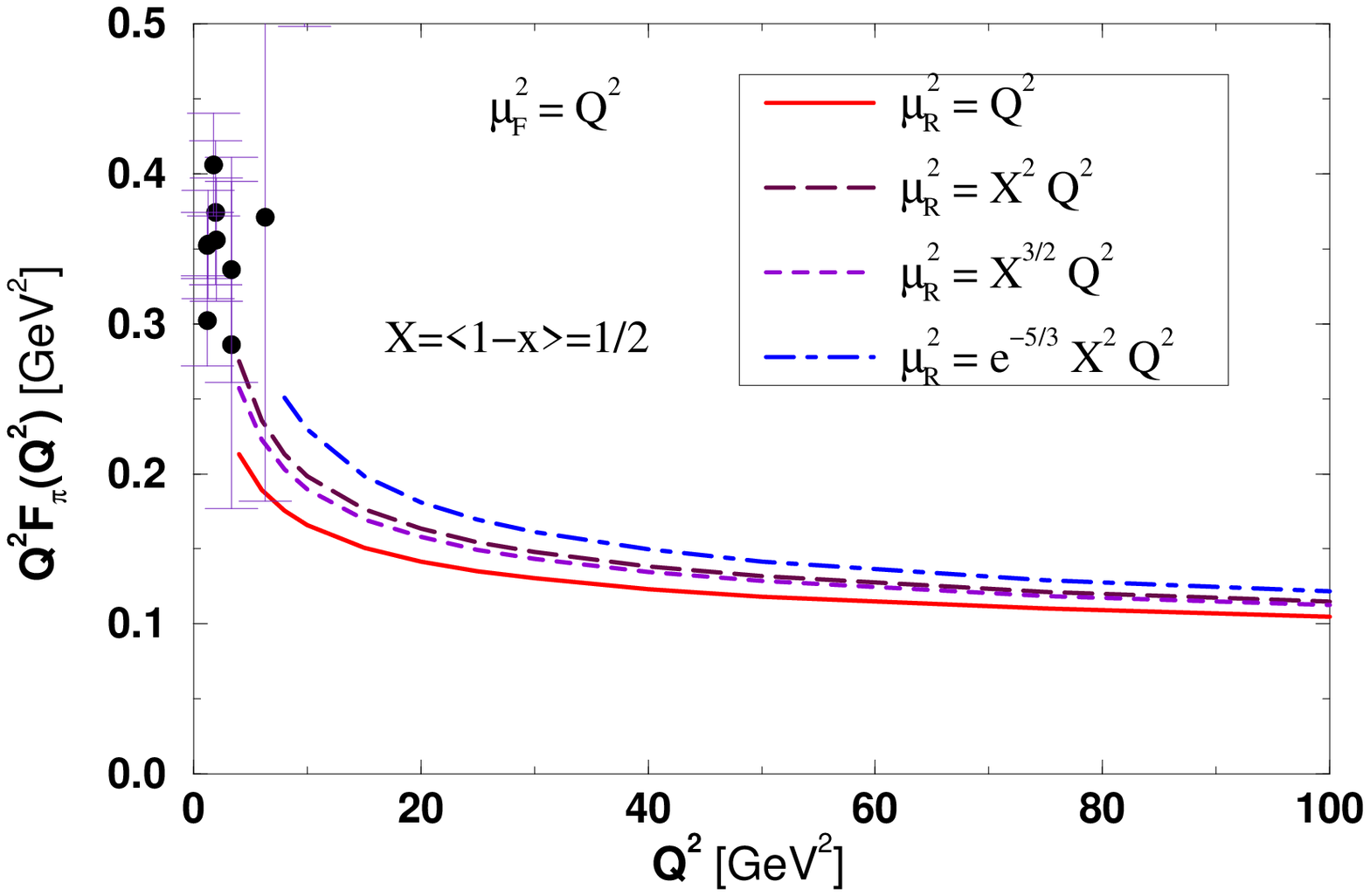,height=4.5cm,width=6.0cm,silent=}
&
\epsfig{file=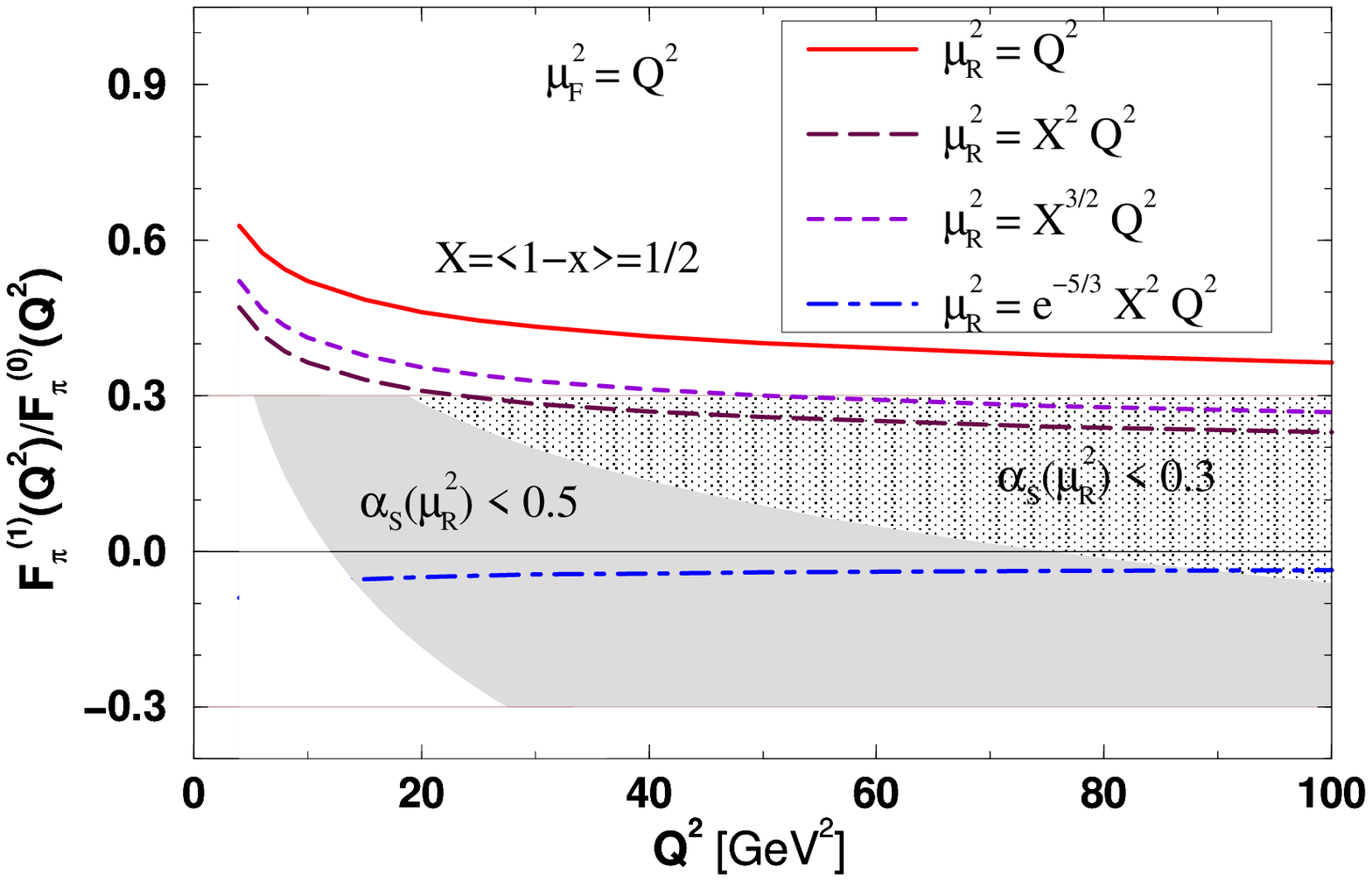,height=4.5cm,width=6.0cm,silent=}
\\
(a) & (b) \\
\end{tabular}
\caption{(a) Leading-twist NLO QCD results for $Q^2 F_{\pi}(Q^2)$ obtained
  using the $\phi_{as}(x,\nIR)$ distribution amplitude and
  the choices of $\nUV$ given by Eqs. \protect\req{eq:nRQ},
  (\protect\ref{eq:nRx2}--\protect\ref{eq:nRx2e}), 
  while $\nIR=Q^2$ and $\left< \x \right>=0.5$.
  The experimental data are taken from \protect\cite{Be78}.
  (b) The ratio $F_{\pi}^{(1)}(Q^2)/F_{\pi}^{(0)}(Q^2)$
  obtained using the same DA and the same choices for
  $\nUV$ and $\nIR$ scales as in Fig. 1a. 
  The shaded area denotes the region of predictions which 
  corresponds to 
  $| F_{\pi}^{(1)}(Q^2)/F_{\pi}^{(0)}(Q^2) |<30\%$, and
  $\alpha_S(\nUV)\!<\!0.5$ ($\alpha_S(\nUV)\!<\!0.3$).
  }
\label{f:ASbA}
\end{figure}

Numerical results of our complete NLO QCD calculation 
for  the pion form factor, $F_{\pi}(Q^2)$,
obtained using the $\phi_{as}(x,\nIR)$ distribution, 
with $\nIR=Q^2$ and different choices for the
renormalization scale $\nUV$ given by \req{eq:nRQ} 
and (\ref{eq:nRx2}--\ref{eq:nRx2e}),
are displayed in Fig. \ref{f:ASbA}a
(in our calculation we take $\Lambda_{\overline{MS}}=0.2$).
The ratio of the NLO to the LO contribution to 
$F_{\pi}(Q^2)$, i.e., $F_{\pi}^{(1)}(Q^2)/F_{\pi}^{(0)}(Q^2)$, 
as a useful measure of the importance of the NLO corrections,
 is plotted as a function of $Q^2$ in Fig. \ref{f:ASbA}b.

The solid curve in Figs. \ref{f:ASbA}a and b 
corresponds to the often encountered choice $\nUV=Q^2$.
The total perturbative prediction $F_{\pi}(Q^2)$ is somewhat below 
the trend indicated
by the presently available experimental data.
To make a meaningful comparison between theory and experiment,
reliable experimental data are needed,
as well as  a reliable estimation and inclusion of 
the soft contributions in a moderate energy region. 
What alarms us is the fact that,
the ratio $F_{\pi}^{(1)}(Q^2)/F_{\pi}^{(0)}(Q^2)$ is rather high and
$F_{\pi}^{(1)}(Q^2)/F_{\pi}^{(0)}(Q^2) \leq 30\%$ is not reached until
$Q^2 \approx 500$ GeV$^2$.
This result seems to question the applicability of the pQCD to exclusive
processes.
The answer to this problem lies in the previously stated inappropriateness
of the choice $\nUV=Q^2$. 
Namely, owing to the partitioning of the overall momentum
transfer $Q^2$ among the particles in the parton subprocess,
the essential virtualities of the particles are smaller than $Q^2$,
so that the ``physical'' renormalization scale, better suited for the
process of interest, is inevitably lower than $Q^2$.

The results 
displayed in Fig. \ref{f:ASbA}a
show that
the total prediction for the pion form factor $Q^2 F_{\pi}(Q^2,\nUV,\nIR)$
depends weakly on the choice of $\nUV$.
This is a reflection of the stabilizing effect that the inclusion of the
NLO corrections has on the LO predictions.
By contrast, 
the results for the ratio
$F_{\pi}^{(1)}(Q^2,\nUV,\nIR)/F_{\pi}^{(0)}(Q^2,\nUV,\nIR)$ 
are sensitive to the choice of $\nUV$.
From the results displayed in Fig. \ref{f:ASbA}b  we find that
by choosing  the renormalization scale 
related to the average virtuality of the particles in the parton
subprocess or given by the BLM scale,
the size of the NLO corrections is significantly reduced and 
reliable predictions
are obtained at considerably lower values of $Q^2$, $Q^2 < 100$ GeV$^2$.

\section{{\it{\Large Conclusions}}}

We conclude by stating that,
regarding the size of the radiative corrections, 
the sHSA can be consistently applied to the calculation
of the pion electromagnetic form factor.
Further investigation of the scale fixing problem \cite{scales} 
as well as the calculation of the NLO prediction in the mHSA 
(the difference should be important only in the region of a few GeV)
remain challenges to future work.

%\newpage

%\acknowledgments
\begin{center}
{\em Acknowledgments}
\end{center}

  This work was supported by the Ministry of Science and Technology
  of the Republic of Croatia under Contract No. 00980102.

\end{document}